\documentclass[graybox, nosecnum]{svmult}

\makeatletter
\providecommand*{\toclevel@author}{0}%
\providecommand*{\toclevel@title}{0}%
\makeatother
\usepackage{newtxtext,newtxmath}
\usepackage{helvet}         
\usepackage{courier}        
\usepackage{type1cm}        
\usepackage{makeidx}         
\usepackage{graphicx}        
\usepackage{multicol}        
\usepackage[bottom]{footmisc}
\usepackage{hyperref}        
\usepackage{soul}            
\hypersetup{colorlinks=true,urlcolor=blue}
\usepackage[square,numbers]{natbib}
\usepackage{multirow}
\usepackage{bm}
\usepackage{amsmath}

\usepackage{amssymb}
\usepackage{cases}

\def\beq{\begin{equation}}
\def\enq{\end{equation}}
\def\beqa{\begin{equation*}}
\def\enqa{\end{equation*}}
\def\bea{\begin{eqnarray}}
\def\ena{\end{eqnarray}}
\def\beaa{\begin{eqnarray*}}
\def\enaa{\end{eqnarray*}}

\makeindex             

\begin{document}
\title*{Fast Radio Bursts}
\author{Di Xiao, Fayin Wang, and Zigao Dai }
\institute{Di Xiao, Fayin Wang \at School of Astronomy and Space Science, Nanjing University, Nanjing 210023, China, \email{dxiao@nju.edu.cn}; \email{fayinwang@nju.edu.cn} \at Key Laboratory of Modern Astronomy and Astrophysics (Nanjing University), Ministry of Education, PR China
\and Zigao Dai \at  Department of Astronomy, University of Science and Technology of China, Hefei 230026, PR China, \email{daizg@ustc.edu.cn}}

%
%
\maketitle
\abstract{The era of fast radio bursts (FRBs) was open in 2007, when a very bright radio pulse of unknown origin was discovered occasionally in the archival data of Parkes Telescope. Over the past fifteen years, this mysterious phenomenon have caught substantial attention among the scientific community and become one of the hottest topic in high-energy astrophysics. The total number of events has a dramatic increase to a few hundred recently, benefiting from new dedicated surveys and improved observational techniques. Our understanding of these bursts has been undergoing a revolutionary growth with observational breakthroughs announced consistently. In this chapter, we will give a comprehensive introduction of FRBs, including the latest progress. Starting from the basics, we will go through population study, inherent physical mechanism, and all the way to the application in cosmology. Plenty of open questions exist right now and there is more surprise to come in this active young field.}
\section{Keywords} 
fast radio burst, neutron star, cosmology
\section{Introduction}
Fast radio bursts (FRBs) are newly-discovered bright millisecond radio transients that flash randomly in the sky. Dating back to 2007, an unknown bright single pulse was found in the archival data of Parkes telescope \cite{Lorimer2007}, marking the beginning of this research field. At first, there is a doubt among the community whether this signal is astrophysical, especially at the moment when ``perytons'' were found \cite{Burke-Spolaor2011}. This query finally came to an end as a new sample of FRBs were discovered in 2013, meanwhile the possibility of terrestrial origins was excluded \cite{Thornton2013}. Since then, scientists showed growing research interest on this kind of bursts and various source models have been proposed. This field became highly active with observational breakthroughs announced consistently in every single year. In 2014 an FRB was found in the archival data of a second telescope other than Parkes \cite{Spitler2014}, further confirming the astrophysical origin. In 2015 Parkes detected a real-time FRB for the first time and multi-wavelength follow-up was triggered \cite{Petroff2015}. In 2016 the first repeating FRB 20121102A was identified \cite{Scholz2016,Spitler2016} and then localized to a dwarf galaxy in the next year \cite{Chatterjee2017,Marcote2017,Tendulkar2017}. This landmark discovery refreshed our knowledge of these bursts and their source models need to be reconsidered. From 2018 Australian Square Kilometre Array Pathfinder (ASKAP) and Canadian Hydrogen Intensity Mapping Experiment (CHIME) started taking data and the total events of FRB accumulated rapidly \cite{Shannon2018,CHIME2019c}. More and more repeating and localized events were reported in the following years \cite{CHIME2019a,CHIME2019b,Bannister2019,Ravi2019,Fonseca2020,Macquart2020,CHIME2021cat}. Things became more interesting in 2020 that the periodic activity of FRB 20180916B was found for the first time \cite{CHIME2020a}. Then a Galactic FRB 20200428A was found to be associated with X-ray bursts from the magnetar SGR 1935+2154 \cite{CHIME2020b,Bochenek2020,Mereghetti2020,Li2021,Ridnaia2021,Tavani2021}, This is such an important milestone that we finally knew the source of these radio bursts. In 2021 several new results were announced, such as the chromatic periodic activity \cite{Pastor-Marazuela2021}, sub-second periodicity \cite{CHIME2021}, highly polarized micro-structure \cite{Nimmo2021a} and unique energy distribution \cite{LiD2021}. With the upgrade of existing instruments and the complete of new facilities, we shall expect more and more important observational progresses coming in the near future.

Synchronized with these observational achievements, theorists have been proposing and optimizing their models continually. Several topics are under hot debate at this moment. First, the classification of FRBs by repeating behaviour is phenomenological. A repeater could be misidentified as a non-repeating event due to selection bias \cite{Palaniswamy2018,Connor2018,Connor2019}. There are dichotomous opinions on whether all FRBs can repeat (maybe just with different repeating modes ?) \cite{Caleb2019}. Second, it remains controversial where the periodic behaviour comes from (orbital motion, precession or spin of the source ?) \cite{Zhang2020}. Moreover, the chromatic active window is hard to understand. Third, the radiation mechanism of these bright radio bursts is an unsolved issue, even for pulsar radio emission \cite{Melrose2017}. Different ways of generating coherence have been discussed, all of which, however, have certain limitations. Actually, there are many more mysteries waiting to be explored in this young research field \cite{Katz2018, Popov2018, Petroff2019,Cordes2019,Zhang2020c,XiaoD2021}. Before a discussion in depth on these topics, we first give a basic introduction on our current understanding of FRBs.

\section{General Properties and Propagation Effects}
The definition of an FRB is descriptive. Basically, a radio burst satisfying three requirements is regarded as an FRB. First, the burst should be really ``fast'', with a duration of milliseconds. Second, the brightness temperature of this burst should be significantly higher than that of pulsar radio emission, which is obtained by 
\bea
T_{\rm B}&\sim& F_\nu d_{\rm A}^2/2\pi k(\nu T)^2 \nonumber\\
&\simeq& 1.1\times10^{35}{\,\rm K}\,\left(\frac{F_\nu}{\rm Jy}\right)\left(\frac{\nu}{\rm GHz}\right)^{-2}\left(\frac{T}{\rm ms}\right)^{-2}\left(\frac{d_{\rm A}}{\rm Gpc}\right)^2,
\label{eq:T_B}
\ena
where $\nu$ is emission frequency, $F_\nu$ is flux density, $d_{\rm A}$ is angular diameter distance, $k$ is Boltzmann constant and $T$ is duration. Typical FRBs have at least $T_{\rm B}>10^{30}\,\rm K$. Third, the total dispersion measure (DM) of this burst is expected to exceed the value that Galactic electrons can contribute in order to distinguish it from a rotation radio transient. However, the last requirement seems loose now since a Galactic FRB was detected. Before reaching the Earth, an FRB signal has experienced multiple propagation effects caused by the plasma on its path.

\subsection{Dispersion}
The dispersion of radio waves by free electrons in the plasma is just similar to visible light spreaded by a prism. It leads to an arrival time delay between high-energy and low-energy photons. Considering two photons with frequencies $\nu_1,\,\nu_2$ ($\nu_1<\nu_2$), the delayed arrival time relates to DM as 
\bea
\Delta t&=&\frac{e^2}{2\pi m_e c}\left(\frac{1}{\nu_{1}^2}-\frac{1}{\nu_{2}^2}\right)\int{\frac{n_e}{1+z}{\rm d}l}\nonumber\\
&\simeq&4.15\,\rm s\left[\left(\frac{\nu_1}{1\,GHz}\right)^{-2}-\left(\frac{\nu_2}{1\,GHz}\right)^{-2}\right]\frac{DM}{10^3\,pc\,cm^{-3}},
\label{eq:timedelay}
\ena
where ${\rm DM}\equiv\int{n_e{\rm d}l/(1+z)}$ is the the electron number density integrated along the traveled
path. Generally, the total DM of an FRB at redshift $z$ consists of four terms,
\bea
{\rm DM=DM_{MW}+DM_{IGM}}+\frac{\rm DM_{host}+DM_{source}}{1+z},
\label{eq:DM_z}
\ena
corresponding to four kinds of intervening plasma from the observer to the burst location, i.e., Milky Way, intergalactic medium, host galaxy and source environment. The Milky way contribution has been well modeled based on the Galactic electron density distribution \cite{Cordes2002,Yao2017}. The second term is of particular interest, the average of which can be written as a function of $z$ \cite{Deng2014},
\bea
\label{eq: DM_IGM}
{\rm \langle DM_{IGM} \rangle }= \frac{3c \Omega_{\rm b}H_0 }{8\pi Gm_p} \int^z_0\frac{F(z')}{E(z^\prime)}{\rm d}z^\prime,
\ena
where $H_0$ is the Hubble constant, $\Omega_{\rm b}$ is the cosmic baryon mass density fraction, $E(z)=H(z)/H_{0}$ and $F(z) \equiv (1+z)f_{\rm IGM}(z)f_{\rm e}(z)$ with $f_{\rm IGM}$ being the fraction of the baryon mass in the IGM. Further, $f_{\rm e}(z) = Y_{\rm H}X_{\rm e,H}(z)+\frac{1}{2}Y_{\rm He}X_{\rm e,He}(z)$, where $Y_{\rm H}$, $Y_{\rm He}$ are the mass fractions of hydrogen and helium, and $X_{\rm e,H}$, $X_{\rm e,He}$ are the corresponding ionization fractions of them, respectively. Due to the close relation between $\rm DM_{IGM}$ and cosmic parameters, Eq. (\ref{eq: DM_IGM}) has been applied widely in cosmology, which will be discussed later.

\subsection{Scattering effect}
Radio photons can be easily scattered by particles on the path, leading to a change of direction. Scattered photons travel a longer way than unscattered ones, and FRB pulse profile may exhibit a tail feature due to different arrival time of photons caused by multi-path propagation. Therefore, the observed pulse width is broadened and we can define a scattering timescale $\tau$. Actually, the observed width $T$ is a combination of intrinsic width $T_{\rm i}$ and several broadening terms \cite{Petroff2019},
\bea
T=\sqrt{T_{\rm i}^2+\tau^2+t_{\rm samp}^2+\Delta t_{\rm DM}^2+\Delta t_{\rm DMerr}^2},
\label{eq:pulse_width}
\ena
where $t_{\rm samp}$ is the data sampling interval. The receiver has finite channel bandwidth $\Delta\nu$, leading to a dispersion smearing term \cite{Cordes2003}
\bea
\Delta t_{\rm DM}=8.3\,\rm\mu s\,DM\Delta\nu_{MHz}\nu_{GHz}^{-3}.
\label{eq:smearing}
\ena
Further, the error of dispersion measure $\rm DM_{err}$ could cause additional smearing \cite{Cordes2003},
\bea
\Delta t_{\rm DMerr}=\Delta t_{\rm DM}(\rm DM_{err}/DM).
\ena
Usually, the long tail of FRB pulse appears as an exponential decay. This can be explained as FRB photons scattered by a thin, extended screen. The scattering timescale is expected to depend sensitively on frequency $\tau\propto\nu^{-4.0}$ in this case. If more realistic scattering medium is assumed, for instance, a Kolmogorov turbulence may change the above dependence to $\tau\propto\nu^{-4.4}$ as long as the minimum turbulence scale is smaller than the diffractive length scale \cite{Xu2016}. 
In practice, $\tau$ is determined by fitting FRB pulse profile on the basis of making assumptions on the intrinsic pulse shape and instrumental broadening. Recent measurements of FRB scattering timescale range from sub-milliseconds to tens of milliseconds. Compared with Galactic pulsars, FRBs look under-scattered in the $\tau$-DM plane \cite{Ravi2019b,Qiu2020}. Astoundingly, a population with large scattering $\tau>10\,\rm ms$ was found in the first catalog of CHIME FRBs, which is difficult to reproduce with current models \cite{CHIME2021cat}.

\subsection{Scintillation}
Similar to the fact that turbulent atmosphere of the Earth leads to the twinkling of stars, turbulent intervening plasma leads to a variation of observed FRB flux density. If the turbulent plasma has a velocity in the direction perpendicular to our line of sight, the interference and diffraction patterns at the position of the observer would vary accordingly. Scintillation has been well studied before in pulsar field  \cite{Rickett1990, Narayan1992}. As the plane waves passing through a turbulent screen, random phase fluctuations are generated. The transverse radius at which the root mean square phase difference is 1 rad can be defined as the diffractive length scale $r_{\rm diff}$. We have two different regimes according to the values of $r_{\rm diff}$ and the Fresnel scale $r_{\rm F}\equiv\sqrt{\lambda D/2\pi}$, where $\lambda$ is wavelength and $D$ is the distance between the screen and the observer. If $r_{\rm diff}>r_{\rm F}$, perturbations to wavefronts are weak and the scintillation timescale is just the Fresnel timescale $t_{\rm scint}\simeq t_{\rm F}=r_{\rm F}/V$, with $V$ being the transverse velocity. This is called the weak scattering regime. In the opposite strong scattering regime $r_{\rm diff}<r_{\rm F}$, diffractive interstellar scintillation is important, giving rise to a scintillation timescale of $t_{\rm scint}\simeq r_{\rm diff}/V$. The scintillation bandwidth $\Delta\nu_{\rm scint}$ is related to scattering timescale by $2\pi\Delta\nu_{\rm scint}\tau=C_1$ where $C_1$ is a constant of order unity \cite{Cordes1998}. Scintillation has been found in several FRB events and it might be responsible for the spectral structure of the Galactic FRB 20200428A \cite{Simard2020}. However, scintillation could hardly be the main reason for the fast variability of light curves in most FRBs \cite{Beniamini2020}. Sometimes, the refractive interstellar scintillation may be relevant on a scale of $r_{\rm ref}=r_{\rm F}^2/r_{\rm diff}$, and the corresponding scintillation timescale is $t_{\rm scint}\simeq r_{\rm ref}/V$, which could be even longer than FRB itself \cite{Scholz2016}. 

\subsection{Plasma lensing} 
Refraction is a common phenomenon occurring when the medium that photons are passing through has a sudden change. Ionized plasma is refractive with an index of refraction $n_{\rm p}=\sqrt{1-\omega_{\rm p}^2/\omega^2}$. From this index we can see that plasma lenses are diverging and highly chromatic \cite{Clegg1998}. If the plasma lens lies exactly on the line of sight, the observed flux will be in the minimum. However, a large magnification could also be achieved if a moderate offset of the lens exists. Different source-lens-observer geometry leads to different time variability of FRB light curves. Practically, we can model plasma lensing in a way similar to gravitational lensing and the major difference is effective deflection potential \cite{Wagner2020}. The deflection angle depends sensitively to the electron number density distribution on the lens plane. Assuming a one-dimension Gaussian lens, the influence of plasma lensing on FRB light curve and spectrum has been discussed \cite{Cordes2017}. Meanwhile, an additional frequency-dependent delay of the arrival time is introduced and this effect should be corrected when inferring DM value \cite{Er2020}. The complex time-frequency pulse structure of FRB 20121102A might be explained by plasma lensing \cite{Hessels2019}. With this effect, both upward and downward frequency drift in a sequence of bursts are expected. However, downward drifting appears more frequently for repeaters in observation \cite{Pleunis2021b}. Note that the downward drifting behaviour has been discussed both in near-field and far-away source models \cite{WangWY2019,Metzger2019}. Furthermore, a relation between sub-burst slope and duration was discovered recently \cite{Chamma2021}, and it was claimed to be well explained by a simple dynamical model \cite{Rajabi2020}.

\subsection{Absorption} 
Radio emission can be absorbed by a few processes, among which the most important is the free-free absorption. Free electrons can gain energy after absorbing photons. For a given electron density, the optical depth can be obtained with the cross section calculated from quantum mechanics. Generally, we can neglect the absorption in the interstellar and intergalactic medium since the number density is usually very low. However, significant absorption can happen if the circumburst medium is dense enough. Models have been proposed that FRB source might reside in a wind nebula \cite{Yang2016,Margalit2018,ZhaoZY2021b}. In this environment, even synchrotron self-absorption may become relevant. The absorbed radio bursts may heat the electrons in the nebula and give rise to the observed persistent radio source (PRS) accompanying FRB 20121102A \cite{LiQC2020}. The other signature of absorption is a low-energy cutoff in FRB spectrum. The current lowest frequency of detection is near 110 MHz for FRB 20180916B \cite{Pastor-Marazuela2021, Pleunis2021a}, indicating that the local environment of this FRB is relatively clean. The lack of bursts below 100 MHz may be due to either absorption or intrinsic FRB radiation spectrum.

\subsection{Faraday rotation}
If the plasma is magnetized, another propagation effect emerges, i.e., the Faraday rotation. For a linearly-polarized electromagnetic (EM) wave, its electric field vector will rotate an angle with respect to the initial direction after passing through this plasma. The observed polarization angle (PA) is 
\bea
\Psi_{\rm obs}(\lambda)=\Psi_0+\frac{e^3\lambda^2}{2\pi m_e^2c^4}\int{n_e(l)B_{\parallel}(l){\rm d}l},
\label{eq:PAobs}
\ena
where $\Psi_0$ is the initial PA, and $B_{\parallel}$ represents the component of the magnetic field along the line of sight. We can define rotation measure as
\bea
{\rm RM}=\frac{e^3}{2\pi m_e^2c^4}\int{n_e(l)B_{\parallel}(l){\rm d}l},
\ena
and for cosmological sources, it is usually scaled as
\bea
{\rm RM\left[\frac{rad}{m^2}\right]}=0.812\int{\frac{n_e[{\rm cm^{-3}}]B_{\parallel}[{\rm \mu G}]}{(1+z)^2}{\rm d}l[\rm pc]}.
\ena
The RM value has a positive sign if the magnetic field points towards us, otherwise it can be negative. We can estimate the RM value by fitting the dependence of $\Psi_{\rm obs}$ on $\lambda$ in Eq. (\ref{eq:PAobs}). In practice, more precise values can be obtained via RMFIT, RM Synthesis or QU-fitting \cite{Hotan2005, Macquart2012, OSullivan2012}. If a same plasma medium dominates the DM and RM value, then the magnetic field strength within this medium can be estimated as \cite{Caleb2018}
\bea
\langle B_{\parallel}\rangle = \frac{{\rm RM}}{0.812 {\rm DM}}\simeq1.232\rm \left(\frac{RM}{rad\,m^{-2}}\right)\left(\frac{DM}{pc\,cm^{-3}}\right)^{-1}\,\mu G.
\ena
The typical RM value is from tens to hundreds of units except for an outlier FRB 20121102A, which has extremely large ${\rm RM}\sim10^5\,\rm rad\,m^{-2}$ and is decreasing with time \cite{Michilli2018,Hilmarsson2021a}. 

The polarization properties are diverse for the whole population. Some FRBs show highly linear polarization , while others are partially polarized or unpolarized. Repeaters like FRB 20121102A and 20180916B are nearly 100\% linearly polarized and their PAs do not vary obviously with time \cite{Gajjar2018,Michilli2018,CHIME2019b,Nimmo2021a}. Non-repeaters usually have different degrees of polarization and their PAs evolve with time \cite{Cho2020, Day2020}. However, no clear dichotomy has been established since a repeater FRB 20180301A showed complex PA swing \cite{Luo2020b}. A recent study found that the polarization degree of repeaters depends on frequency, which can be explained by multi-path scatter \cite{FengY2022}. This is a reminder that the observed polarization is not the polarization at source. Therefore, it remains questionable whether intrinsic polarization property can be used to classify FRBs.  

\section{Global Statistical Properties and Population Study}

With the increase of FRB event number, it is possible to study the statistical properties of them. Till now the total number has exceeded six hundred on Transient Name Server \footnote{https://www.wis-tns.org/}, of which 24 are found to repeat and 19 are well localized. Many statistical studies have been done and put strong constraints on source models. Besides, a comparison of FRBs with other short-duration transients such as magnetar outbursts, giant pulses and solar radio bursts have been carried out \cite{WangFY2017,ChengYJ2020,Lyu2021,WangFY2021}.

\subsection{Energy, pulse width and waiting time distribution}
The burst energy can be calculated as
\bea
\label{eq:ene}
E = \frac{4 \pi d_{\rm L}^2 F \nu_{\rm c}}{1+z},
\ena
where $d_{\rm L}$ is luminosity distance and $F$ is burst fluence. Here the central frequency $\nu_{\rm c}$ is used instead of the bandwidth of the receiver $\Delta\nu$, which is reasonable if the emission extends beyond $\Delta\nu$ \cite{ZhangB2018}. However, some FRBs are so narrow-banded that the bandwidth of the burst itself should be used in energy estimation \cite{Aggarwal2021}. Early studies showed that the energy distribution of FRB 20121102A followed a power-law form \cite{Law2017,Gourdji2019,WangFY2019}. Recently, a bimodal energy distribution of this FRB was found thanks to the low energy threshold of FAST telescope \cite{LiD2021}. 

The width distribution looks distinct for different burst samples. A power-law distribution was found for FRB 20121102A using GBT data \cite{ChengYJ2020}, while a log-normal fit was adopted for the FAST sample \cite{LiD2021}. Detailed analysis of CHIME FRB catalog suggested that the width distribution seems different between one-off events and repeaters. Repeating FRBs tend to have longer duration and more complex morphology (sub-burst structures) \cite{Pleunis2021b}. The intrinsic width is directly related to the radiation mechanism and it remains unclear what dependence it has.

The waiting time distribution is an important factor characterizing repeating FRBs. A Possionian distribution for FRB 20121102A has been disfavored \cite{WangFY2017,Oppermann2018,Cruces2021}, implying that bursts are unlikely to occur from a pure stochastic process. Intriguingly, double peaks showed up in the waiting time distribution for the FAST sample. One is around several milliseconds and the other is around tens of seconds \cite{LiD2021}. Besides, the second peak was found to be consistent with Poisson statistics and this was confirmed by Arecibo observation soon later \cite{Jahns2022}. It is worth further discussion whether two types of burst process exist.

\subsection{Host galaxy properties}
At the moment 19 FRBs have been precisely localized with identified host galaxies. The detailed information of these galaxies can be found in FRB host database \footnote{http://frbhosts.org}. It is obvious that there is a diversity among host galaxies and their stellar mass spans a wide range. Most FRB hosts are moderately star-forming galaxies and lie offset from the star-forming main sequence. Also, a dearth of red galaxies was found \cite{Bhandari2022}. Substantial analysis showed FRB hosts do not track stellar mass or star formation rate (SFR). The population is statistically consistent with hosts of short gamma-ray bursts (SGRBs) and core-collapse supernovae (CCSNe) \cite{Bhandari2022}. Some FRBs have large offsets from galaxy center that consistent with SGRBs, indicating that they might originate from similar events like binary neutron star (NS) mergers \cite{WangFY2020}. No clear difference in host galaxy properties for repeating and one-off FRBs has been found yet. All the above results need further justification with a larger sample of FRB hosts in the future.

\subsection{Luminosity function and redshift evolution}
The all-sky rate of FRB is as high as $10^3\sim10^4\,\rm sky^{-1}\,day^{-1}$ given the current telescope sensitivity. This rate also depends on the observing frequency, as most FRBs are discovered in L-band and CHIME observing band (400-800 MHz). Basically, the event rate density is more relevant for characterizing transients. However, most FRBs except 19 localized events do not have credible distance measurements, therefore the redshift evolution $\Psi(z)$ is highly uncertain. The errorbar in the estimation of the cosmological volumetric rate can be reduced with the accumulation of localized events. Beforehand, method of constraining luminosity function and redshift evolution using statistical properties has been proposed \cite{Bera2016, Caleb2016, Macquart2018a, Macquart2018b, Niino2018}.  

It is unknown whether the luminosity function evolves with redshift, and usually a non-evolutionary luminosity function $\Phi(L_\nu)$ is assumed just for simplicity. Therefore a separable function can be defined $\Theta(L_\nu,z)\equiv\Phi(L_\nu)\Psi(z)$ to incorporate both factors. Different $\Theta(L_\nu,z)$ can be constructed if special forms of $\Phi(L_\nu)$ and $\Psi(z)$ are adopted. The commonly-used forms for $\Phi(L_\nu)$ include standard candle, power-law, log-normal and Schechter functions \cite{Caleb2016, Niino2018, Fialkov2018, Luo2018}. Meanwhile, the most natural model for redshift evolution is that $\Psi(z)$ traces the cosmic SFR \cite{Caleb2016, Niino2018}, or a delayed SFR model if FRBs are produced by compact mergers \cite{ZhangGQ2019, Locatelli2019}. Also, the possibility that $\Psi(z)$ traces cosmic stellar mass density has been discussed \cite{Niino2018}. With a constructed $\Theta(L_\nu,z)$, it is possible to predict the distributions of some observational properties and compare with the real data \cite{Macquart2018b}. Feasible observational properties include the cumulative distribution of FRB flux density/fluence \cite{Macquart2018a,Patel2018,James2019}, value of $\langle V/V_{\max}\rangle$ test \cite{Shannon2018, Locatelli2019}, DM distribution \cite{Deng2019,ZhangGQ2019,Bhattacharya2019,Lu2019,Lu2020a} and sensitivity-dependent detection rate \cite{Lawrence2017,Bhandari2018}. Note that there is a degeneracy of $\Theta(L_\nu,z)$ with FRB spectral index \cite{Niino2018}, hence a Bayesian method is recommended \cite{Lu2019,Luo2020}. 

\subsection{FRB classification}
Now we have hundreds of FRBs and the sample is enlarging very quickly. Individual FRBs look distinct in many ways such as pulse morphology, spectro-temporal behavior and polarization property. A fundamental question is now under hot debate: are there two or more populations of FRBs and how can they be classified?

A straightforward classification is via repetition behavior. FRBs are deemed repeaters if a second burst is detected, otherwise are deemed one-off events. This classification method was favored by the evidence that repeaters and one-off events seem to be different in a few aspects such as pulse morphology and spectral property \cite{Pleunis2021b,ZhongSQ2022}. However, this result is not conclusive enough using current FRB sample. Moreover, this criterion is phenomenological and suffers from selection bias. If the subsequent burst is beyond the observing band or is much dimmer than the original burst, the probability of missing it in observation is quite high \cite{Connor2018,Palaniswamy2018}. In this case, a real repeater is misidentified as a non-repeating one. In a conservative way, maybe all FRBs can repeat just with different repeating modes \cite{Caleb2019}. Several studies have explored this possibility and it has been proposed that the number fraction of repeaters can judge \cite{Ai2021,Gardenier2021a}. The conservative scenario will be verified if this fraction approaches unity with accumulating observing time in the future. Otherwise, this fraction will peak at a certain time, then the classification by repetition is non-trivial.

Alternatively, it is worthwhile to look for more physical criteria of classification. One feasible candidate is brightness temperature \cite{XiaoD2022a}. Besides FRBs, there are many kinds of short-duration radio transients such as pulsar radio emission, giant pulses and nanoshots. They cluster in different regions on the spectral luminosity-duration phase space plot, and the main difference between them is brightness temperature \cite{Nimmo2021b}. The large FAST sample of FRB 20121102A has been classified in this way and even a two-parameter correlation for the classified bursts was found \cite{XiaoD2022a}. This has been further confirmed by a classification of FRBs in CHIME catalog \cite{Chaikova2022}.

\subsection{Periodicity}
The periodic activity of repeaters is an interesting phenomenon discovered in 2020 \cite{CHIME2020a}. The period of FRB 20180916B is about 16 days, and was further found to be chromatic. The active window arrives earlier and looks narrower at higher frequencies \cite{Pastor-Marazuela2021}. Except for this event, a plausible period of $\sim160$ days has been claimed for FRB 20121102A \cite{Rajwade2020b, Cruces2021}. Searches for periodicity of other frequent repeaters like FRB 20190520B and FRB 20201124A have reported null result \cite{XuH2021,NiuCH2021}. More intriguingly, sub-second periodicity has been found for FRB 20191221A with a significance of 6.5 sigma and for FRB 20210206A and 20210213A with lower confidence levels \cite{CHIME2021}. It remains largely unknown what causes periodic behaviour and how common it is among all repeaters.

There are a few studies dedicated to solve this problem and three kinds of origin exist in current models. However, these models can not explain all the observational properties \cite{WeiYJ2022}. The first possible origin is orbital motion. The FRB-emitting NS is in a binary system and the companion could be an O/B type stars or a compact object \cite{Lyutikov2020a, ZhangXF2020,Gu2020,Ioka2020,Zhang2020b}. FRBs can be observed once these two objects are in a certain orbital phase. Recently the chromatic active window has been reproduced in a Be/X-ray binary scenario \cite{LiQC2021}. The second option is that periodicity is due to NS precession, with both free and forced precession being discussed \cite{Levin2020, YangH2020, Zanazzi2020, Nikolaevich2020}. Forced precession could be caused by the surrounding disk \cite{Chen2020,Tong2020}. The third possibility is that the rotational period of NS could be relevant \cite{Beniamini2020}. This model might have advantages in explaining the sub-second periodicity, in turn, is unlikely responsible for days-long period. There is no evidence for the existence of such a slow-rotating NS and it can hardly produce FRBs even if there exists. 

\section{Physical Mechanism of FRBs}
FRBs are diverse in so many aspects that it is not easy to figure out what are the sources and by which mechanism they are produced. As a comparison, the radiation mechanism for pulsar radio emission is still under debate after half a century, and the FRB radiation mechanism seems more complicated. In consideration of the much higher brightness temperature of FRBs, the physical conditions of the NS are generally more extreme. This is the core issue of FRB field and we might have a long way to go. However, the accumulating observational data have been giving us clues consistently (see \cite{Zhang2020c} for a review). 

\subsection{Radiation mechanism}
The brightness temperature of FRBs is so high that the emission must be coherent. In general, there are three ways  to generate coherence in astrophysics \cite{Melrose2017}. The first one is antenna mechanism. Charged particles form a bunch and they emit in a same phase almost simultaneously. The emitted power is in proportion to the square of particle number \cite{Benford1977}. The second way is maser. Somehow population inversion is realized and there is available free energy leading to negative absorption \cite{Hoshino1991}. The third option is relativistic
plasma emission by reactive instabilities \cite{Melrose1991}. The particles' kinetic energy is initially transferred to Langmuir waves through a streaming instability and finally converted to escaping radio emission. All three ways above have been discussed extensively for pulsar radio emission and the first two have been applied to FRBs in detail.

\subsubsection{Antenna mechanism}
Coherent curvature radiation by bunches is a close-in mechanism that happens inside the magnetosphere of a NS. Let us consider an electron moving along a field line with a curvature radius $\rho$, the characteristic angular frequency and radiation spectrum is  
\bea
\omega_c&=&\frac{3}{2}\gamma^3\frac{c}{\rho},\nonumber\\
\frac{{\rm d}P}{{\rm d}\omega}&=&\frac{\sqrt{3}e^2\gamma}{2\pi \rho}\frac{\omega}{\omega_c}\int_{\omega/\omega_c}^{\infty}{K_{5/3}(y){\rm d}y},
\label{eq:curva}
\ena
where $\gamma$ is the Lorentz factor of this electron and $P$ is the integrated power over solid angle. Once a bunch of $N_{\rm e}$ electrons is formed, coherence can be generated if the emitted wave phases of individual electrons are almost the same. The total radiation intensity of this bunch is 
\bea
\frac{{\rm d}E_{\rm tot}}{{\rm d}\omega {\rm d}\Omega}=\frac{e^2\omega^2}{4\pi^2c}\left |\int_{-\infty}^{+\infty}\sum_j^{N_{\rm e}}\bm{n}\times(\bm{n}\times\bm{\beta}_j)e^{i\omega(t-\bm{n}\cdot\bm{r}_j(t)/c)}{\rm d}t\right |^2,
\label{eq:bunch}
\ena
where $\bm{n}$ is the unit vector pointing to the observer, and $\bm{\beta}_j$, $\bm{r}_j$ are the velocity, position vector of the $j$-th electron respectively. Since the emission is highly beamed, Eq. (\ref{eq:bunch}) can be approximated by
\bea
\frac{{\rm d}E_{\rm tot}}{{\rm d}\omega {\rm d}\Omega}&\simeq&\frac{e^2\omega^2}{4\pi^2c}\left|\int_{-\infty}^{+\infty}\bm{n}\times(\bm{n}\times\bm{\beta})e^{i\omega(t-\bm{n}\cdot\bm{r}(t)/c)}{\rm d}t\right|^2\nonumber\\
&\,&\times\left|\sum_j^{N_{\rm e}}e^{-i\omega(\bm{n}\cdot\Delta\bm{r}_j/c)}\right|^2.
\label{eq:bunch2}
\ena
where $\bm{r}(t)$ is the position vector of the first electron and $\Delta\bm{r}_j(t)\equiv\bm{r}_j(t)-\bm{r}(t)$ is the relative displacement of the $j$-th electron. Defining a frequency $\omega_L\equiv2c/(L\cos\theta)$ with $\theta$ being the observing angle and $L$ being the bunch length, then the phase stacking term
\bea
F_\omega\equiv\left|\sum_j^{N_{\rm e}}e^{-i\omega(\bm{n}\cdot\Delta\bm{r}_j/c)}\right|^2\simeq\begin{cases}
	N_{\rm e}^2, &\omega\ll\omega_L,\\
	N_{\rm e}^2\left(\frac{\omega}{\omega_L}\right)^{-2}, &\omega_L\ll\omega\ll\omega_{\rm coh}.
\end{cases}
\label{eq:phasefactor}
\ena
where $\omega_{\rm coh}\sim(\rho/L)^2\omega_L$ is the maximum frequency for coherence \cite{Yang2018}. Therefore, the emission power of a bunch is usually taken to be proportional to $N_{\rm e}^2$ for simplicity. The detailed radiation spectrum for a three-dimensional bunch filled with electrons of power-law energy distribution has been calculated analytically \cite{Yang2018}. Some commonly-used expressions for estimating the total luminosity are listed below. If Eq. (\ref{eq:curva}) is integrated over $\omega$, the curvature radiation power of a single electron is
\beq
P_{\rm curv}=\frac{2}{3}\frac{\gamma^4e^2c}{\rho^2}\simeq4.61\times10^{-15}\,{\rm erg\,s^{-1}}\gamma_{2.5}^4\rho_8^{-2}.
\label{eq:Pcurv}
\enq
The emission of different bunches is generally incoherent \cite{Yang2018}, thus the total luminosity $L_{\rm curv}$ is proportional to the number of bunches $N_{\rm b}$. We can express it as
\beq
L_{\rm curv}=N_{\rm b}N_{\rm e}^2\gamma^2P_{\rm curv}.
\label{eq:Lcurv}
\enq
This mechanism works inside the NS magnetosphere, in which the typical electron number density is characterized by Goldreich-Julian (GJ) density \cite{Goldreich1969}
\bea
n_{\rm GJ}=\Omega B/(2\pi e c)=6.94\times10^7\,{\rm cm^{-3}}B_{{\rm s},15}P^{-1}R_8^{-3}, 
\label{eq:nGJ}
\ena
where a magnetar engine with surface field strength $B_{\rm s}\sim10^{15}\,\rm Gauss$, rotational period $P\sim 1\,\rm s$ and emission radius $R\sim10^8\,\rm cm$ is assumed. Usually a multiplicity factor $\mathcal{M}$ is needed. Coherence requires the length of the bunch being smaller than the emission wavelength $\lambda$. Meanwhile, the transverse size of causally connection is $\sim\gamma\lambda$ \cite{Kumar2017}. Therefore the number of electrons in one bunch is approximated as 
\bea
N_{\rm e}\simeq\mathcal{M}n_{\rm GJ}\pi\gamma^2\lambda^3=5.89\times10^{17}\mathcal{M}\gamma_{2.5}^2\nu_{\rm GHz}^{-3}B_{{\rm s},15}P^{-1}R_8^{-3},
\label{eq:Ne}
\ena
and Eq. (\ref{eq:Lcurv}) turns into 
\bea
L_{\rm curv}\simeq1.60\times10^{31}{\,\rm erg\,s^{-1}}N_{{\rm b},5}\mathcal{M}^2\gamma_{2.5}^{10}\nu_{\rm GHz}^{-6}\rho_8^{-2}B_{{\rm s},15}^2P^{-2}R_8^{-6}.
\label{eq:Lcurvvalue}
\ena
This expression is useful for the rough estimation on the properties of the source. It is worth mentioning that coherent inverse Compton scattering (ICS) has been proposed as a possible mechanism for FRBs \cite{Zhang2022}. In this case the emission power of a single electron is enhanced by several orders of magnitude and the total luminosity can be much higher than the value in Eq. (\ref{eq:Lcurvvalue}).

The applicability of antenna mechanism for FRBs has been widely discussed and a few models based on it have been proposed \cite{Dai2016,Kumar2020,Lu2020b,WangWY2020}. Since FRBs are regarded to originate inside the magnetosphere, the diverse PA swing, nano-second variability and frequency drift can be explained by this mechanism \cite{Luo2020b,Lu2022,WangWY2019}. However, a long-standing problem of this mechanism is the formation and maintenance of these bunches \cite{Saggion1975,Cheng1977, Kaganovich2010}. Further, the coherent emission is suppressed if the density of surrounding plasma is too high \cite{Gil2004}. It will take a while to solve these leftover problems from pulsar field. 

\subsubsection{Synchrotron maser emission from magnetized shocks }
The other plausible mechanism that has caught substantial attention is synchrotron maser emission. It occurs as a shock propagating through a magnetized medium. Basically, charged particles behind the shock front can achieve bunching in gyration phase. Particle-in-cell (PIC) simulation suggested that the distribution of these particles in momentum space shows a cold ``ring'' structure \cite{Alsop1988, Gallant1992, Amato2006, Plotnikov2019}, which can be expressed as 
\bea
f(u_\perp, u_\parallel)=\frac{1}{2 \pi u_{0}} \delta(u_\perp-u_{0})
\delta(u_\parallel),
\label{eq:ring}
\ena
where $u$ is four-velocity and symbols $\parallel$, $\perp$ are with respect to the magnetic field direction. This implies that population inversion is reached and synchrotron maser instability can develop \cite{Alsop1988}. The dispersion relation leads to two unstable branches corresponding to electromagnetic and magnetosonic waves. The growth of the former can be very effective therefore coherent emission is produced \cite{Hoshino1991, Amato2006}. This signal precedes the emission of shock-heated particles in the downstream, thus is called an EM precursor. Traditionally in the conservation equations of energy and momentum for the shock jump condition, this EM precursor is unexpected for ideal MHD plasma \cite{Gallant1992}. However, PIC simulations show that additional wave fluctuations exist and could dissipate the flow energy \cite{Amato2006,Plotnikov2019}. The fraction of energy carried away by EM precursor $f_\xi$ can be expressed using the fluctuation parameter $\xi$ of upstream field  \cite{Gallant1992}, 
\bea
f_\xi&\equiv&\frac{\xi}{1+1/\sigma}\frac{1-\beta_{\rm shock}}{1+\beta_{\rm shock}},
\ena
where $\sigma$ is the magnetization parameter of the upstream medium and $\beta_{\rm shock}$ is the shock velocity. Simulations show that this fraction peaks at $\sigma\sim0.1$ with a value of $\sim10\%$, and has an asymptotic form
of $f_\xi\simeq7\times10^{-4}/\sigma^2$ for the case of $\sigma\gg1$ \cite{Plotnikov2019}. Meanwhile, the radiation spectrum has a peak at $\omega_{\rm peak}\simeq 3\omega_{\rm p}\max[1,\,\sqrt{\sigma}]$ in the post-shock frame, where $\omega_{\rm p}$ is plasma frequency. Overall, the spectrum is irregular and narrow-banded with $\Delta\omega/\omega_{\rm peak}\sim$ a few \cite{Plotnikov2019}.

A big advantage of this mechanism is that it is the only process supported by first-principle calculations (PIC simulations). It should work in various situations within the whole universe. Its application in FRBs has been discussed and the flat PA curve, downward drifting of repeaters are naturally expected \cite{Metzger2019}. Besides, the predicted high-energy counterpart has been observed for Galactic FRB 20200428A \cite{Mereghetti2020,Li2021,Ridnaia2021,Tavani2021}. Another simulation showed that low-amplitude Alfvén waves from a magnetar quake can be convert to plasmoids, afterwards, collision with the wind will lead to blast waves \cite{YuanYJ2020}. However, this mechanism also have some defects as being disfavored recently by observations, such as the baryonic mass budget \cite{WuQ2020b}, the PA swing \cite{Luo2020b} and nano-second variability \cite{Nimmo2021a}.  

To conclude this section, there are far more than two mechanisms proposed for FRBs, many of which was reinvented from pulsar field \cite{Lyutikov2021a}. At the moment, none of them seems perfect since distinct burst morphology and weird spectral structures have not been explained well. It is unclear whether multiple mechanisms can work for FRBs. The debate on radiation mechanism may last a while, just similar to the situation of pulsar and GRB field. 
\subsection{Source models}
The source of FRBs has been discussed extensively in literature. The number of proposed models exceeds that of FRB events once for a time \cite{Platts2019}. The first extinction of models occurs with the discovery of the repeating FRB 20121102A, since many models are catastrophic and only viable for one-off events. Later on, the range of source models has been further narrowed down as the Galactic FRB 20200428A was found to originate from a magnetar \cite{CHIME2020b,Bochenek2020}. There is an evidence that magnetars can produce all the observed FRBs from population synthesis \cite{Gardenier2021b}. However, it is still early to conclude that magnetars do it all. Multiple sources may be responsible for different FRBs, and different predictions will be tested by future observations.

Source models can be classified as ``close-in'' and ``far-away'' depending on whether the site of emission region is inside or outside the magnetosphere. Based on coherent curvature radiation, a detailed close-in model has been developed and is introduced here \cite{Kumar2020,Lu2020b}. The crustal quake of a magnetar will launch Alfv{\'e}n waves moving along the field lines. These waves propagate outward in the polar cap region where plasma density decreases with radius. Charge starvation occurs beyond a critical radius where the density is too low to support the electric current. A parallel electric field will develop and electrons, positrons get accelerated. Bunches can form due to the counter-streaming of pairs and coherent emission is produced. Assuming an amplitude $\delta B$ and wavelength component perpendicular to the magnetic field $\lambda_\perp$ of the Alfv{\'e}n waves, the critical density for charge starvation as a function of radius $R$ is 
\bea
n_{\rm c}(R)=(10^{16}\,{\rm cm^{-3}})\frac{\delta B_{11}}{\lambda_{\perp,4}}\left(\frac{R_\ast}{R}\right)^3,
\ena
where $R_\ast$ is the magnetar radius. The isotropic FRB luminosity is then 
\bea
L_{\rm iso}\simeq\frac{16e^2cR^5\gamma^2n_{\rm c}^2l_\parallel}{3\rho^2},
\ena
where $l_\parallel$ is the bunch length. This model has been applied to FRB 20200428A and the spectra can be explained \cite{Yang2020b}. Note that in this model the X-ray bursts should arrive earlier than the FRB \cite{Lu2020b}, and radio emission needs to break out from the X-ray fireballs \cite{Ioka2020b}.  

The other well-developed model is far-away and based on sychrontron maser emission mechanism \cite{Metzger2019,Beloborodov2020}. A magnetar born from a CCSN has a strong wind, and a nebula could form due to wind interacting with SN ejecta. This active magnetar produces flares irregularly and the flare ejecta may collide with the wind or the leftover ion shell from the previous flare, leading to strong shocks. Both the wind and ion shell are magnetized, however, with different $\sigma$ values. Assuming that the upstream medium has a density profile of $n_{\rm ext}\propto r^{-k}$, the dynamical evolution of the system is similar to that of GRB afterglow \cite{Sari1995}. The Lorentz factor of the shocked region $\Gamma$ evolves with time $t$ as
\bea
\Gamma  \propto  \left\{
\begin{array}{lr}
	t^{\frac{(k-2)}{2(4-k)}}, &\,\, t\ll\delta t  \\
	t^{\frac{(k-3)}{2(4-k)}}, &\,\, t\gg\delta t
\end{array}
\right. ,
\label{eq:Gammaevo}
\ena
where $\delta t$ denotes the crossing time of reverse shock. The intrinsic spectrum of synchrotron maser emission peaks at $\nu_{\rm pk} \approx \Gamma \omega_{\rm peak}/(2\pi)$ in lab frame. However, the induce Compton scattering could be important and the observed peak frequency $\nu_{\rm max}$ is higher than this value. The time evolution of $\nu_{\rm max}$ is 
\bea
\nu_{\rm max} \propto \nu_{\rm pk}^{5/4}t^{1/4} \propto \left\{
\begin{array}{lr}
	t^{-\frac{2+7k}{4(8-2k)}}, & \,\,\,\,t \lesssim \delta t,  \\
	t^{-\frac{2k+7}{4(8-2k)}}, & \,\,\,\,t \gtrsim \delta t. \\
\end{array}
\right.
\label{eq:numax1}
\ena
Therefore the downward drifting can be explained. Meanwhile, the fraction of flare energy that goes into FRB emission is further reduced to $f\sim10^{-6}-10^{-5}\ll f_\xi$ \citep{Metzger2019,XiaoD2020}, thus
\bea
L_{\rm maser}\sim f E_{\rm flare}/T\sim 10^{37}\,{\rm erg\,s^{-1}}f_{-6}E_{{\rm flare},40}T_{\rm ms}.
\label{eq:Lmaser}
\ena
where the flare energy $E_{\rm flare}$ is scaled with the typical X-ray bursts energy \cite{Gogus1999,Gogus2000}. Giant flares with $E_{\rm flare}>10^{43}\,\rm erg$ is needed to produce high-luminosity FRBs. This model has also been applied to FRB 20200428A and the observed radio to X-ray ratio $E_{\rm radio}/E_X\sim10^{-5}$ matches model prediction \cite{Margalit2020, XiaoD2020}. Further refinement of this model can be realized if more realistic upstream medium is assumed \cite{Beloborodov2020,XiaoD2020}. Also, simulations show that such relativistic shocks can be produced from magnetar quakes \cite{YuanYJ2020}. 

Except for these two, there are many other variants of magnetar models \cite{Dai2016,Wadiasingh2019,Wadiasingh2020,Lyutikov2020b,Lyubarsky2020}. The mystery of source has not been completely solved and it is possible that multiple sources can produce FRBs. Besides, it remains unclear whether repeaters and one-off events have the same origin. This points to FRB classification problem again and at this time it is hard to tell whether a genuinely one-off FRB exists. This can be verified once an association with a cataclysmic event is observed in the future.

\subsection{FRB counterpart}
Up to now, only two kinds of EM signal are found to accompany FRBs. One is PRS for FRB 20121102A and 20190520B \cite{Chatterjee2017,NiuCH2021}, and the other is X-ray bursts for FRB 20200428A \cite{Mereghetti2020,Li2021,Ridnaia2021,Tavani2021}. The counterparts of FRBs are closely related to sources, and different models have predicted dozens of them. For instance, one source model suggested that FRBs can be produced from the inspiral of two NSs \cite{WangJS2016}. We can naturally expect gravitational wave (GW) signal following FRBs and various GW counterparts may also be there such as GRBs and kilonovae. However, these counterparts are usually very faint and the event needs to be close enough for them to be detected. Right now only two extragalactic FRBs are within the detection horizon of current LIGO/Virgo, i.e., FRB 20181030A from a star-forming spiral galaxy NGC 3252 \cite{Bhardwaj2021b} and FRB 20200120E from a globular cluster in M81 \cite{Kirsten2022}. The rarity of nearby FRBs may be the main reason for the absence of multi-wavelength and multi-messenger counterparts. 

Nevertheless, it is meaningful to search for counterparts for the current FRB sample. Generally, there are three kinds of strategies. First one is the direct rapid follow-up of luminous FRB events \cite{Nunez2021}. The second is searching for an association in archival data. The aim is to find any transient consistent with an FRB in sky position and occurrence time. Many kinds of coincident transients can be searched such as SNe, kilonovae and high-energy bursts. Also, the association with GWs and high-energy neutrinos can be tested. The third strategy is monitoring some special repeaters regularly and hope for a good luck. Unfortunately, most of the observational campaigns designed to search for FRB counterparts has returned null results, however, some upper-limits can be given \cite{Kilpatrick2021}. The other reason for non-detection may be that the optical counterparts can be as short as FRBs \cite{Yang2019}, therefore high-cadence observation is needed for future searches \cite{Tingay2021}. 

\section{Applications in Cosmology}
\label{sec5}
The cosmological origin and precise DM measurements of FRBs make them an attractive cosmological probe \cite{XiaoD2021}. The dispersion of FRBs accounts for every single
ionized baryon along the line of sight. Therefore, they can be used to study the baryonic matter of the Universe \cite{McQuinn2014,Macquart2020}, including the amount and locations. 

A more ambitious goal is to use the DM-$z$ relation to measure the proper distance the Universe \cite{Yu2017}, the equation of state of dark energy \cite{Zhou2014,Walters2018,Zhao2020,QiuXW2022}, Hubble constant \cite{Li2018,Hagstotz2022,WuQ2021}, Hubble parameter \cite{WuQ2020a}, dark matter \cite{MunozJ2016,WangYK2018} and the cosmic Helium and hydrogen reionization history \cite{Zheng2014,Zhangz2021}. 
This would require at least two conditions. First, a large sample of FRBs should be localized with measured distances. This can be achieved by future Square Kilometre Array (SKA). Second, different DM contributions should be separated well. According to Eq. (\ref{eq:DM_z}), the cosmological information containing in DM$_{\rm IGM}$ is attractive. However, the inhomogeneity of IGM has a significant effect on the DM-$z$ relation. Meanwhile, the DM contributed by host galaxies is hard to be determined from observations now.

\begin{figure}
    \centering
    \includegraphics[width = 0.8\textwidth]{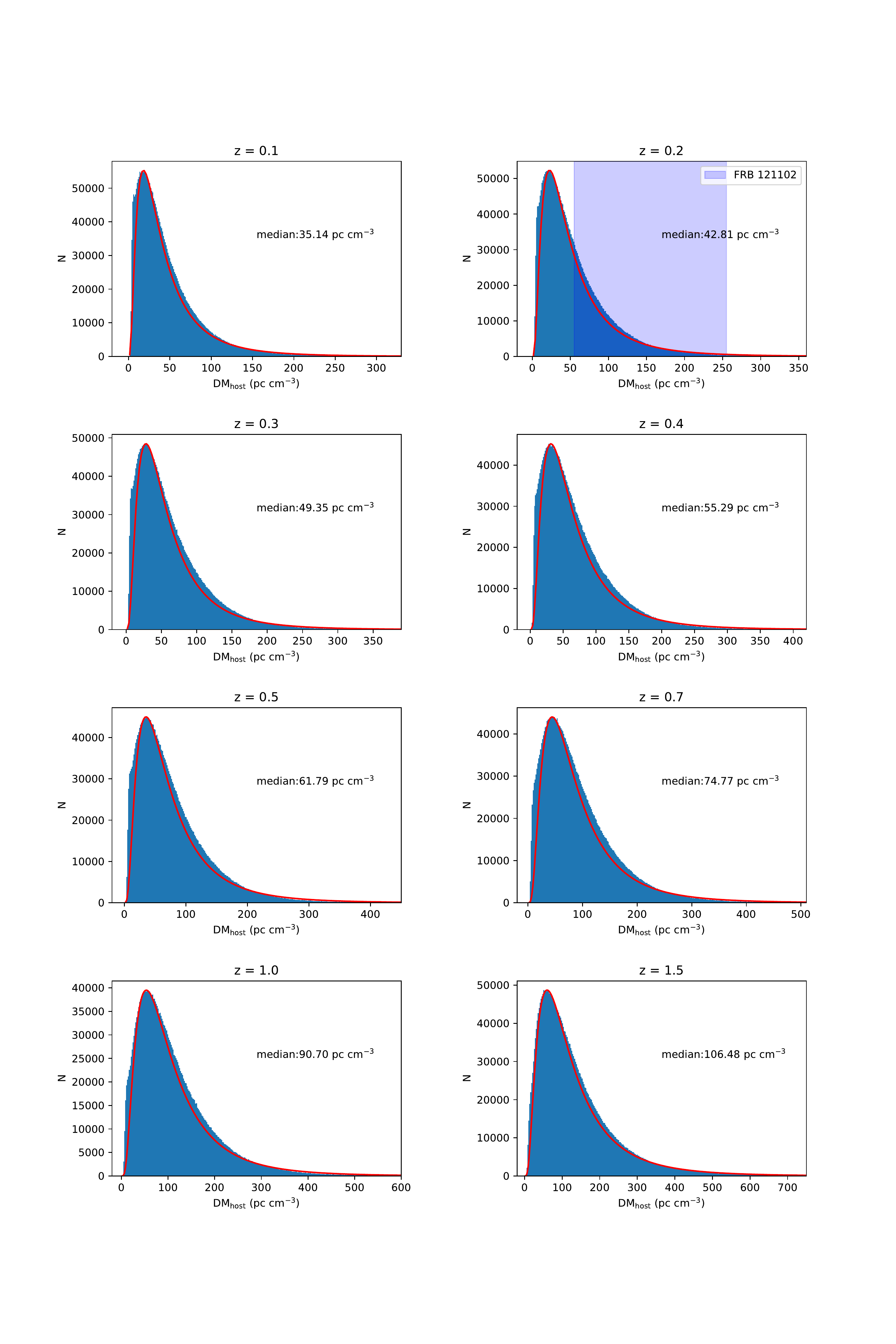}
    \caption{The distributions of $\mathrm{DM}_{\rm{host}}$ at different redshifts for repeaters like FRB 20121102A. }
    \label{fig:Disrepeat}
\end{figure}

\subsection{{\rm DM} contribution of host galaxy and source environment}
The DMs of FRBs could contain significant contributions from host
galaxies or local environment of FRBs. The best way to estimate the
DM contribution of host galaxies is using cosmological simulations.
By choosing a large sample of galaxies from the IllustrisTNG simulation \cite{Springel2018}, 
the distributions of DM$_{\rm host}$ were derived to have a long tail, indicating some light paths pass through the whole galaxy. A log-normal function could be used to fit the DM$ _{\mathrm{host}} $ distribution \cite{ZhangGQ2020},
\begin{equation}
	\label{eq:lognorm}
	P(x; \mu, \sigma) = \frac{1}{x\sigma \sqrt{2\pi}} \exp\left(- \frac{(\ln x - \mu)^2}{2\sigma^2}\right),
\end{equation}
The mean and variance of this distribution
are $ e^\mu $ and $ e^{(2\mu+\sigma^2)}[e^{\sigma^2} - 1] $, respectively. In Figure  \ref{fig:Disrepeat},
the red lines are the best fits and the histograms are derived from the IllustrisTNG simulation.

For the source contribution, it may depend on specific origin models. The association of FRB 20200428A with the Galactic SGR 1935+2154 indicates that at least some FRBs are powered by magnetars. It is generally believed that magnetars could form via CCSNe and compact binary mergers. Since the ejecta mass of compact mergers is usually very low, the corresponding contribution DM$_{\rm source}$ is negligible  \cite{WangFY2020,ZhaoZY2021a}. However, for the core-collapse case, this DM$_{\rm source}$ could be large \cite{Piro2016,Yang2017}. The observation of FRB 20190520B with the largest host contribution supports this scenario \cite{NiuCH2021,Katz2022,ZhaoZY2021b}.

\begin{figure}
    \centering
    \includegraphics[width = 0.8\textwidth]{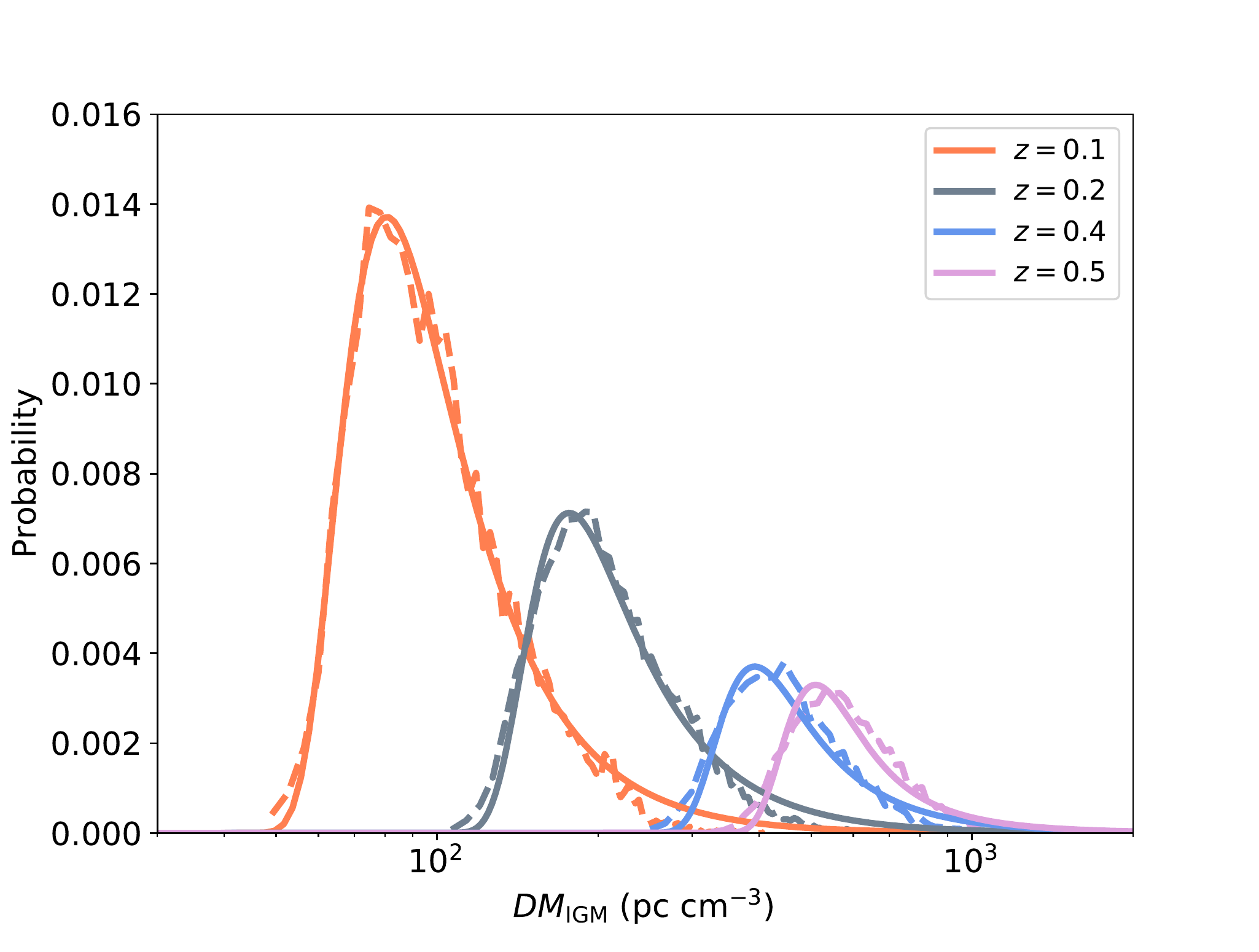}
    \caption{The distributions of $\mathrm{DM}_{\rm{IGM}}$ at different redshifts (dash lines) from the IllustrisTNG simulation. The solid lines are the best fits using Eq. (\ref{PIGM}).}
    \label{fig:igmdis}
\end{figure}

\subsection{Fluctuations in IGM}

The free electrons along different sight-lines in IGM is not uniform, so it is hard to determine the true value of $\rm DM_{IGM}$. 
A quasi-Gaussian function was proposed to fit the distribution of $\rm DM_{IGM}$ \cite{McQuinn2014}, which includes random variations in the electron distribution.
This function can be expressed as 
\begin{equation}\label{PIGM}
P_{\rm IGM}(\Delta) = A \Delta^{-\beta} \exp  \left[-\frac{(\Delta^{-\alpha}-C_0)}{2\alpha^2\sigma^2_{\rm DM}} \right], \Delta > 0,
\end{equation}
where $\Delta \equiv \rm DM_{IGM}/\langle DM_{IGM}\rangle$. Two parameters are adopted as $\alpha =3$ and $\beta = 3$. 
From Figure \ref{fig:igmdis} we can see that this model provides a good fit of those derived from cosmological simulations \cite{Zhangz2021}. 
Therefore, in order to properly handle the DM-$z$ relation of FRBs, the distributions of DM$_{\rm host}$ and DM$_{\rm IGM}$ should be considered carefully \cite{Macquart2020,WuQ2021}. Especially, taking DM$_{\rm host}$ as a constant or Gaussian distributed is not reliable.

\begin{figure}
    \centering
    \includegraphics[width = 0.8\textwidth]{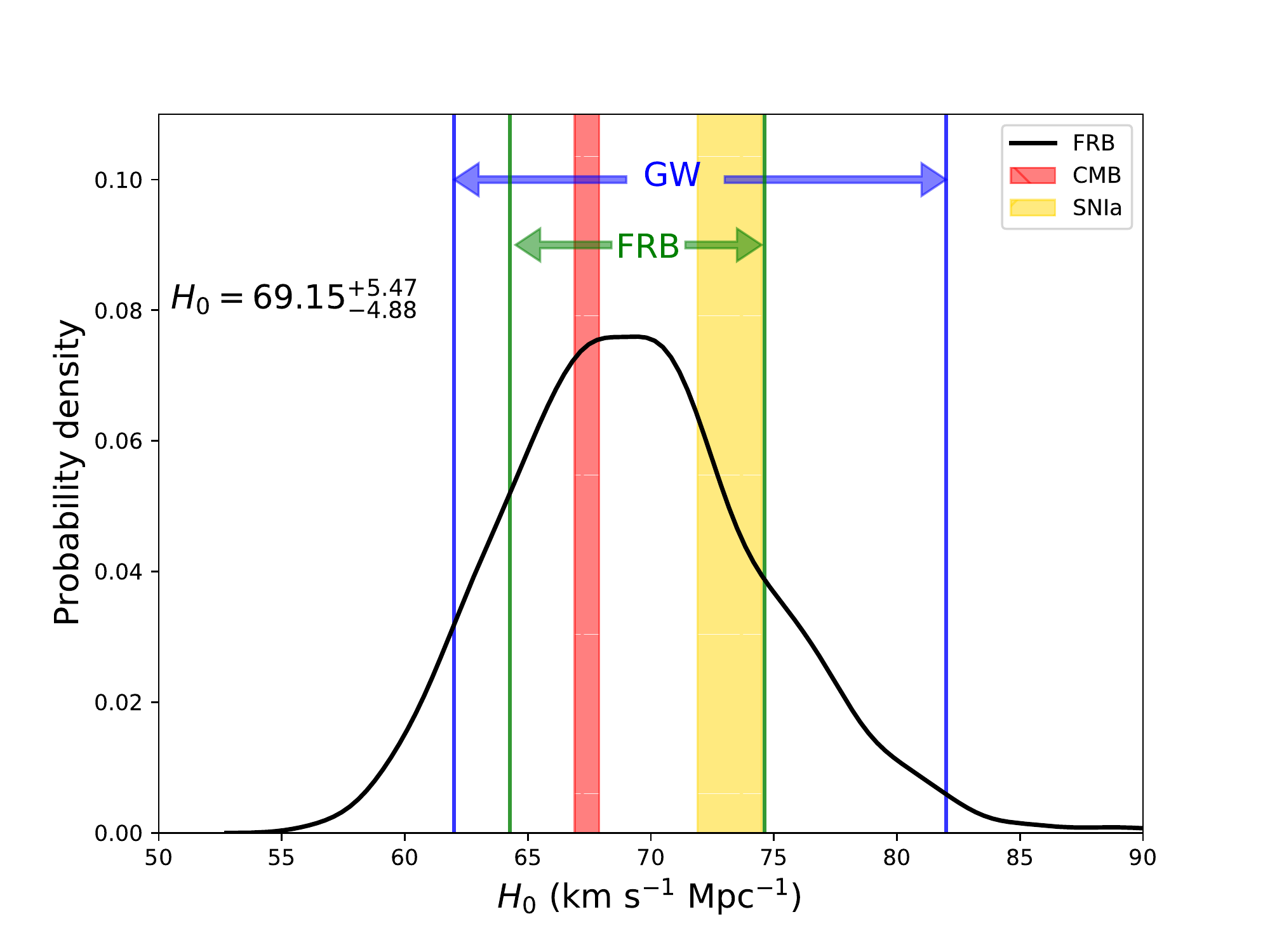}
    \caption{The Hubble constant $H_0$ measured by 18 localized FRBs. The $H_0$ value derived from cosmic microwave background, type Ia supernovae and GW170817 are also given.}
    \label{fig:H0}
\end{figure}

Figure \ref{fig:H0} shows the Hubble constant $H_0$ derived from 18 localized FRBs after taking the above distributions of DM$_{\rm host}$ and DM$_{\rm IGM}$. The best-fitting value is $H_0=69.15^{+5.47}_{-4.88}$ km/s/Mpc \cite{WuQ2021}, which is consistent with those derived from cosmic microwave background \cite{PlanckCollaboration2020} and type Ia supernovae \cite{Riess2021} at 1$\sigma$ confidence level.

\section{Future Prospect}
The research field of FRBs is very young and undergoing rapid development right now. A lot of mysteries and weird features need to be explored in the future. We summarize three main categories of open questions here.

The first thing is the source and radiation mechanism. Although a magnetar was found to be capable of producing FRBs, it remains questionable whether all magnetars can do so. What is the key required physical condition? Can other objects or systems produce FRBs as well? Does repeaters and one-off events have the same origin? Is is possible that one-off FRBs originate from some catastrophic events (like NS-NS mergers)? The observed total energy of FRBs spans a wide range, are they produced by a single mechanism or multiple coherent mechanisms? Where is the emission site, inside or outside the magnetosphere? Finally, what kind of observation is needed to settle these debates?

The second aspect is about the population study. Can all FRBs repeat? What makes the difference in repeating modes? Are there any other physical criteria of classification (e.g., brightness temperature or inherent polarization)? Is the periodic behaviour common among all repeaters? Why does the periodicity vary from sub-seconds to tens of days? How can we narrow down the luminosity function and redshift evolution jointly using different observations? Where is the FRB location in the host galaxy and what is the typical ambient environment? What dominates the diversity of FRB observational properties, inherent physics or propagation effects?  

Last but not the least, it would be very helpful to study FRBs in a multi-messenger point of view. Therefore, searching for FRB counterparts is a timely and meaningful approach. Till now only two kinds of counterparts have been identified, i.e., the PRS of FRB 20121102A, 20190520B and the X-ray bursts accompanying FRB 20200428A. Are they ubiquitous for other FRBs but currently unobservable? If not, what are the physical requirements and radiation mechanisms for them? Are there any other kinds of counterparts, for instance, multi-wavelength ``afterglows'' similar to GRBs? Is it possible to find an association with GW events or high-energy neutrinos? All of the above questions need further investigation and hopefully some answers can be found in the next decade.

\acknowledgement
This work was supported by the National Key Research and Development Program of China (Grant No. 2017YFA0402600),  the National SKA Program of China (grant No. 2020SKA0120300), and the National Natural Science Foundation of China (Grant Nos. 11833003, U1831207, 11903018).


\bibliographystyle{spbasic} %
\bibliography{FRBlatest} 

\end{document}